\documentclass[aps,prl,twocolumn,showpacs,amsmath,amssymb,amsfonts,nofootinbib,long]{revtex4}

\usepackage{graphicx}

\begin{document}


\title{The fields of uniformly accelerated charges in de Sitter spacetime}

\author{Ji\v{r}\'{\i} Bi\v{c}\'ak}
\email{bicak@mbox.troja.mff.cuni.cz}

\author{Pavel Krtou\v{s}}
\email{Pavel.Krtous@mff.cuni.cz}

\affiliation{
  Institute of Theoretical Physics,
  Faculty of Mathematics and Physics, Charles University,\\
  V Hole\v{s}ovi\v{c}k\'{a}ch 2, 180 00 Prague 8, Czech Republic
  }

\date{July 1, 2002} 

\begin{abstract}
The scalar and electromagnetic fields of charges uniformly
accelerated in de~Sitter spacetime are constructed. They
represent the generalization of the Born solutions
describing fields of two particles with hyperbolic motion
in flat spacetime. In the limit $\Lambda\rightarrow 0$, the
Born solutions are retrieved. Since in the de~Sitter
universe the infinities $\scry^\pm$ are spacelike, the
radiative properties of the fields depend on the way in
which a given point of $\scry^\pm$ is approached. The
fields must involve both retarded and advanced effects:
Purely retarded fields do not satisfy the constraints at
the past infinity $\scry^-$.
\end{abstract}

\pacs{04.20.-q, 04.40.-b, 98.80.Hw, 03.50.-z}

\maketitle


The question of the electromagnetic field and associated radiation from
uniformly accelerated charges has been one of the best known \vague{perpetual
problems} in classical physics from the beginning of the past century. In the
pioneering work in 1909, Born gave the time-symmetric solution for the field of
two point particles with opposite charges, uniformly accelerated in opposite
directions in Minkowski space. In the 1920s Sommerfeld\-, von~Laue, Pauli,
Schott and others discussed the properties of the field. The controversial
point that the field exhibits radiative features but that the radiation
reaction force vanishes for the hyperbolic motion, and related questions, was
discussed in many articles from 1960s onward. Even the December 2000 issue of
Annals of Physics contains three papers \cite{EriksenGron:2000} with numerous
references on \vague{electrodynamics of hyperbolically accelerated charges}.

In general relativity, solutions of Einstein's equations, representing
\vague{uniformly accelerated particles or black holes}, are the
\emph{only} explicitly known exact \emph{radiative} spacetimes
describing \emph{finite} sources. They are asymptotically flat at null
infinity \cite{BicakSchmidt:1989} (except for some special points) and
have been used in gravitational radiation theory, quantum gravity and
numerical relativity (cf. review \cite{Bicak:Ehlers}). One of the best
known examples is the $C$-metric, describing uniformly accelerated black
holes. There exists also the $C$-metric for a nonvanishing cosmological
constant~$\Lambda$. However, no general framework is available to
analyze these spacetimes for $\Lambda\neq 0$ as that given in
Ref.~\cite{BicakSchmidt:1989} for $\Lambda=0$.

In this Letter, we present the generalization of the Born solutions for
scalar and electromagnetic fields to the case of two charges uniformly
accelerated in de~Sitter universe, and explicitly show how in the limit
${\Lambda\rightarrow0}$ the Born solutions are retrieved. We also study
the asymptotic expansions of the fields in the neighborhood of future
infinity $\scry^+$. In de~Sitter spacetime, conformal infinities,
$\scry^\pm$, are \emph{spacelike}, which implies the presence of
particle and event horizons. It is known \cite{PenroseRindler:book}
that the radiation field is \vague{less invariantly} defined when
$\scry^+$ is spacelike (it depends on the direction in which $\scry^+$
is approached), but no explicit model appears to be available so far.
Our solutions can serve as prototypes for studying these issues.

\begin{figure}
\selectlcform 
{  
\vspace*{41.5pt}
}{ 
}  
\includegraphics{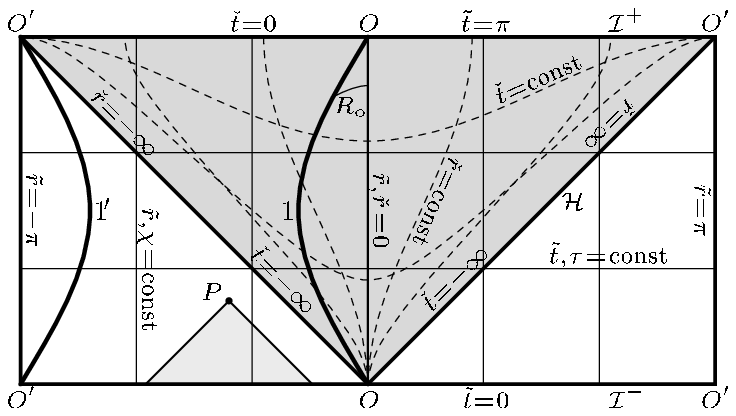}
\caption{\label{fig:dS}
The conformal diagram of de~Sitter spacetime.
\selectlcform 
{  
The lines of coordinates $\{\tlt,\tlr\}$ and $\{\hct,\hcr\}$ are shown.
Uniformly
}{ 
Uniformly
}  
accelerated particles
move along worldlines $1$ and $1'$. The shaded region is the domain of
influence of $1$, its boundary $\mathcal{H}$ is the \vague{creation light cone}
of this particle \vague{born} at ${\tlt=0}$ at \vague{point} $O$. Retarded
fields of $1$ and $1'$ cannot affect point $P$; a Coulomb-type field, however,
cannot vanish there.
}%
\end{figure}

In recent work~\cite{BicakKrtous:asds}, we analyzed fields of accelerated
sources to show the \emph{insufficiency of purely retarded fields in de~Sitter
spacetime}. Consider a point $P$ near $\scry^-$ whose past null cone will not
cross the particles' worldlines (Fig.~\ref{fig:dS}). The field at $P$ should
vanish if an incoming field is absent. However, the \vague{Coulomb-type} field
of particles cannot vanish there because of Gauss law \cite{Penrose:1967}. The
requirement that the field be purely retarded leads, in general, to a bad
behavior of the field along the \vague{creation light cone} of the
\vague{point} at which a source enters the universe (see
Ref.~\cite{BicakKrtous:asds} for detailed discussion).

It is natural to use de~Sitter space for studying radiating sources in
spacetimes which are not asymptotically flat and possess spacelike infinities:
It is the space of constant curvature, conformal to Minkowski space, and with%
\selectlcform 
{  
Huygens principle (no tails of radiation)%
}{ 
Huygens principle%
}  
satisfied for conformally invariant fields.
The de~Sitter universe also plays an important role in cosmology --- not only
in the context of inflationary theories but also as the \vague{asymptotic
state} of standard cosmological models with $\Lambda>0$, which has been indeed
suggested by recent observations. In addition, the Born fields generalized to
de~Sitter space should be relevant from quantum perspectives: for example, for
studying particle production in strong fields, or accelerating detectors in the
presence of a cosmological horizon.

The de~Sitter universe has topology ${S^3\times\realn}$. The metric in
standard \vague{spherical}
\selectlcform 
{  
coordinates\footnote{
  It is convenient to allow the angular coordinates $\chi$, $\tht$,
  $\ph$ on $S^3$ to attain values in $\realn$ and use the
  identifications
  ${\chi\protect\cong\chi+2\pi}$,
  ${\tht\protect\cong\tht+2\pi}$,
  ${\ph\protect\cong\ph+2\pi}$,
  ${\{\chi,\tht,\ph\}} \protect\cong {\{-\chi,\pi-\tht,\ph+\pi\}}$,
  and
  ${\{\chi,\tht,\ph\}} \protect\cong {\{\chi,-\tht,\ph+\pi\}}$.
  Thus, the \vague{radial} coordinate $\chi$ can be negative,
  the points with ${\chi<0}$ being identical to those with
  ${\abs{\chi}>0}$, located \vague{symmetrically} with respect to
  the origin ${\chi=0}$. The same convention is used
  for $\tlr$, $\nlr$, $\hcr$, and $R$ (see Appendix in
  \cite{BicakKrtous:asds} for details).}
${\{\tau,\chi,\tht,\ph\}}$ is%
}{ 
coordinates (note~\cite{note:negr}) is%
}  
\begin{equation}\label{dSMtrctauchi}
  \mtrc_\dS = -\grad\tau\formsq +
  \alpha^2 \cosh^2(\tau/\alpha)\;
  \bigl(\grad\chi\formsq
  +\sin^2\chi\;\sphmtrc\bigr)\commae
\end{equation}
where
  $\sphmtrc = \grad\tht\formsq + \sin^2\!\tht\,\grad\ph\formsq$,
  ${\tau\in\realn}$, and $\alpha^2={3}/{\Lambda}$.
\selectlcform 
{  
Putting
\begin{equation*}
  \chi=\tlr\comma
  \tau = \alpha\log\Bigl(\tan\frac\tlt2\Bigr)\comma
  \tlt\in\langle0,\pi\rangle
\end{equation*}
in Eq.~\eqref{dSMtrctauchi}, the de~Sitter metric can be written in the form
\begin{equation}\label{dSMtrctl}
  \mtrc_\dS = \frac{\alpha^2}{\sin^2\tlt}\;\bigl(
  -\grad\tlt\,\formsq+\grad\tlr\formsq+
  \sin^2\!\tlr\;\sphmtrc\bigr)\period
\end{equation}
The corresponding coordinate lines are drawn in the Penrose diagram of
de~Sitter space in Fig.~\ref{fig:dS}. The lines ${\tlr=\pi}$ and
${\tlr=-\pi}$ are identified, the spacelike hypersurfaces ${\tlt=0}$
and ${\tlt=\pi}$ represent past ($\scry^-$) and future
($\scry^+$) infinities of de~Sitter spacetime.%
}{ 
Putting
  $\chi=\tlr$, $\tau = \alpha\log\tan(\tlt/2)$,
  $\tlt\in\langle0,\pi\rangle$,
in Eq.~\eqref{dSMtrctauchi}, the de~Sitter metric can be written in the form
\begin{equation}\label{dSMtrctl}
  \mtrc_\dS = \alpha^2\sin^{-2}\tlt\;\bigl(
  -\grad\tlt\,\formsq+\grad\tlr\formsq+
  \sin^2\!\tlr\;\sphmtrc\bigr)\period
\end{equation}
The lines ${\tlr=\pi}$ and ${\tlr=-\pi}$ are identified, the spacelike
hypersurfaces ${\tlt=0,\pi}$ represent $\scry^-$ and $\scry^+$
(Fig.~\ref{fig:dS}).
}  

By employing conformal techniques, we recently studied \cite{BicakKrtous:asds}
two particles moving with%
\selectlcform 
{  
uniform acceleration\footnote{
  ``A uniform acceleration'' means that the (proper) time
  derivative $\dot{a}^\alpha$ of the acceleration, projected into the
  hypersurface orthogonal to the four-velocity, vanishes.
  The magnitude of the acceleration is constant.}
in de~Sitter space.
The worldlines of both uniformly accelerated particles
are given by
\begin{equation*}
\begin{gathered}
\tan\tlt =
-\frac{\cosh{\beta_\oix}}{\sinh\bigl(
\frac{\lambda_\dS}{\alpha}\cosh{\beta_\oix}\bigr)}
\commae\\
\tan\tlr =
\pm\frac{\sinh{\beta_\oix}}{\cosh\bigl(
\frac{\lambda_\dS}{\alpha}\cosh{\beta_\oix}\bigr)}
\spcpnct;
\end{gathered}
\end{equation*}
they are plotted as $1$ and $1'$ in Fig.~\ref{fig:dS}.%
}{ 
uniform acceleration (note \cite{note:constacc}) in de~Sitter space. Their
worldlines are plotted in Fig.~\ref{fig:dS} as $1$, $1'$ (for explicit formulae
see Ref.~\cite{BicakKrtous:asds}, Eq.~(4.4); see also Eqs.~\eqref{AccMondSsym},
\eqref{AccMonMink} below).%
}  
Both particles start at antipodes of the spatial section of de~Sitter space at
$\scry^-$ and move one towards the other until ${\tlt=\pi/2}$, the moment of
the maximal contraction of de~Sitter space. Then they move, in a
time-symmetric manner, apart from each other until they reach future infinity
at the antipodes from which they started. Their physical velocities, as measured in the
\vague{comoving} coordinates%
\selectlcform 
{  
${\{\tau, \chi, \tht, \ph\}}$ of Eq.~\eqref{dSMtrctauchi},
have simple forms
\begin{equation*}
v_\chi=\sqrt{\mtrc_{\chi\chi}}\;\frac{d\chi}{d\tau}=
\mp \frac{a_\oix\alpha\tanh(\tau/\alpha)}{
\sqrt{1+a_\oix^{2}\alpha^2\tanh^2(\tau/\alpha)}}\commae
\end{equation*}%
}{ 
${\{\tau,\chi,\tht,\ph\}}$,
have simple forms
  ${v_\chi=\sqrt{\mtrc_{\chi\chi}}\,{d\chi}/{d\tau}}=
  {\mp a_\oix\alpha\tanh(\tau/\alpha)\,
  [1+a_\oix^{2}\alpha^2\tanh^2(\tau/\alpha)]^{-1/2}}$,%
}  
where $\abs{a_\oix}$ is the magnitude of their acceleration. In contrast to the
flat space case, the particles do not approach the velocity of light in the
\vague{natural} global coordinate system. They are causally disconnected
(Fig.~\ref{fig:dS}) as in the flat space case:
No signal from one particle can reach the other particle.%

\begin{figure}
\includegraphics{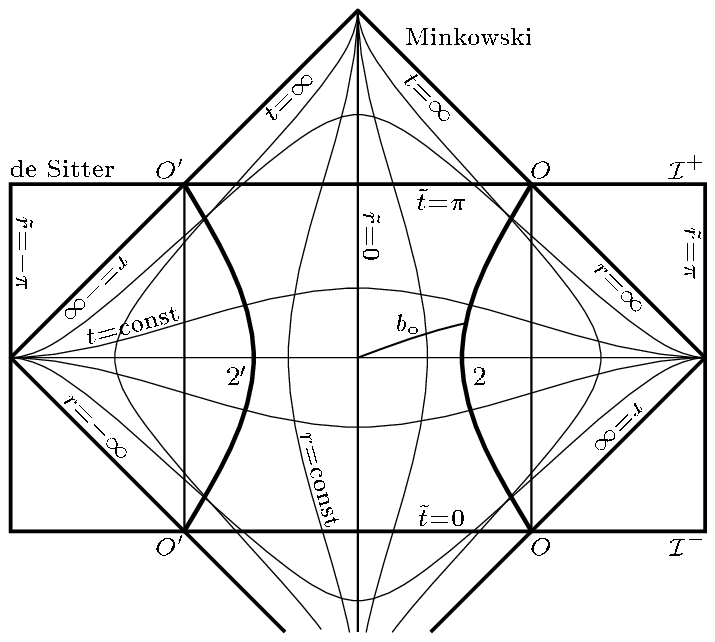}
\caption{\label{fig:dSMink}
The worldlines $2$, $2'$ of uniformly
accelerated charges symmetrically located with respect to the origins of
both de~Sitter and conformally related
Minkowski spacetimes.}
\selectlcform 
{  
\vspace*{20pt}
}{ 
}  
\end{figure}

Two charges moving along the orbits of the boost Killing vector in flat space
are \emph{at rest} in the Rindler coordinate system and have a constant
distance from the spacetime origin, as measured along the slices orthogonal to
the Killing vector. Similarly, the worldlines $1$ and $1'$ are the orbits of
the \vague{static} Killing vector ${\cvectil{T}}$ of
de~Sitter space.%
\selectlcform 
{  
Introducing static coordinates
  ${T\in\realn}$, ${R\in\langle0,\alpha\rangle}$
($\tht$, $\ph$ unchanged) by
\begin{equation*}
  T = \frac\alpha2
  \log\frac{\cos\tlr-\cos\tlt}{\cos\tlr+\cos\tlt}
  \comma
  R=\alpha\frac{\sin\tlr}{\sin\tlt}\commae
\end{equation*}
in which the de~Sitter metric becomes
\begin{equation*}
  \mtrc_\dS =
  -\biggl(1-\frac{R^2}{\alpha^2}\biggr)\grad T\formsq
  +\biggl(1-\frac{R^2}{\alpha^2}\biggr)^{-1}\grad R\formsq
  +R^2\;\sphmtrc\commae
\end{equation*}
we find that the uniformly accelerated particles $1$, $1'$
have the four-acceleration $-(R_\oix/\alpha^2)\,\cvectil{R}$, the magnitude $\abs{a_\oix}$
being constant, and they are at rest at
\begin{equation*}
  R = \pm R_\oix =
  \mp\frac{a_\oix\alpha^2}{\sqrt{1+a_\oix^2\alpha^2}}
  \period
\end{equation*}%
}{ 
In static coordinates
  ${\{T,R,\tht,\ph\}}$,
  $T = {\frac\alpha2 \log[({\cos\tlr\mspace{-0.5mu}-\mspace{-0.5mu}\cos\tlt\mspace{1.5mu}})/({\cos\tlr\mspace{-0.5mu}+\mspace{-0.5mu}\cos\tlt\mspace{1.5mu}}})]$,
  $R = {\alpha\, {\sin\tlr}/{\sin\tlt}}$,
the particles $1$, $1'$ are at rest at
  ${R \mspace{-2mu}=\mspace{-2.5mu} \pm R_\oix \mspace{-2.5mu}=\mspace{-2.5mu} \mp{a_\oix\alpha^2}/{\sqrt{1\!+\!a_\oix^2\alpha^2}}}$,
with four-accelerations
  $-(R_\oix/\alpha^2)\,\cvectil{R}$.%
}  
The particle $1$ ($1'$) has, as measured at fixed $T$, a constant proper
distance from the origin ${\tlt=\pi/2}$, ${\tlr=0}$ (${\tlr=\pi}$). As with
Rindler coordinates in Minkowski space, the static coordinates cover only a
\vague{half} of de~Sitter space; in the other half the Killing vector
$\cvectil{T}$ becomes spacelike.

By the conformal transformation of the boosted Cou\-lomb fields in Minkowski
space, we constructed~\cite{BicakKrtous:asds} test scalar and electromagnetic
fields produced by charges moving along the worldlines $1$, $1'$ in de~Sitter
space. The scalar field from two \emph{identical} scalar charges $\SFq$ is
given~by%
\selectlcform 
{  
\begin{gather}
  \SF_\sym = \frac{\SFq}{4\pi}\frac1\Kfact
  \commae\label{SFsym}\\
  \frac{\Kfact}{\alpha} =
  \Bigl[\Bigl(\sqrt{1+a_\oix^{2}\alpha^2}
  + a_\oix R \cos\tht\Bigr)^2
  \!- \Bigl(1-\frac{R^2}{\alpha^2}\Bigr)\Bigr]^{\frac12}
  \commae\label{Kfact}
\end{gather}
}{ 
\begin{gather}
  \SF_\sym = (\SFq/{4\pi})\;\Kfact^{-1}
  \commae\label{SFsym}\displaybreak[0]\\
  \Kfact =
  \bigl[\alpha^2\bigl(\sqrt{1+a_\oix^{2}\alpha^2}
  + a_\oix R \cos\tht\bigr)^2
  - \alpha^2+R^2\bigr]^{\frac12}
  \label{Kfact}
\end{gather}%
}  
(Ref.~\cite{BicakKrtous:asds}, Eq.~(5.4)), whereas the electromagnetic field
due to \emph{opposite} charges $+\EMq$ and $-\EMq$ is
(Ref.~\cite{BicakKrtous:asds}, Eq.~(5.7))
\begin{gather}\label{EMsymtl}
\begin{split}
  \EMF_\sym &= -\frac{\EMq}{4\pi}\,
  \frac1{\Kfact^3}
  \frac{a_\oix\,\alpha^4}{\sin^3\tlt}
  \;\Bigl[
  \cos\tlt\sin^2\tlr\sin\tht\,
  \grad\tlr\wedge\grad\tht
  \\&\mspace{12mu}
  +(a_\oix^{-1}\sqrt{a_\oix^{2}+\alpha^{-2}}\sin\tlr +
  \sin\tlt\cos\tht)\,
  \grad\tlt\wedge\grad\tlr
  \\&\mspace{12mu}
  -\sin\tlt\cos\tlr\sin\tlr\sin\tht\,
  \grad\tlt\wedge\grad\tht
  \Bigr]\period
\end{split}
\end{gather}
We call these smooth (outside the sources) fields symmetric because they can be
written as a symmetric combination of retarded and advanced effects from both
charges.

Although Eqs.~\eqref{SFsym} and \eqref{EMsymtl} represent fields due to
uniformly accelerated charges in de~Sitter space, their relation to the
Born solutions is not
\selectlcform 
{  
transparent. In order
to arrive at such a relation we have to consider sources located
symmetrically with respect to the origin ${\tlr=0}$, similarly as the
charges in the Born solution move along hyperbolae symmetrical with respect
to the origin in Minkowski space. Hence, instead of the worldlines $1$
and $1'$ we have to consider the worldlines
$2$ and $2'$ (Fig.~\ref{fig:dSMink})%
}{ 
because the sources are not located
symmetrically with respect to ${\tlr=0}$. Hence, we consider the
worldlines $2$ and $2'$ (Fig.~\ref{fig:dSMink})%
}  
which, due to homogeneity and isotropy of de~Sitter space, also represent
uniformly accelerated particles. These worldlines and the resulting fields can
be obtained from Eqs.~\eqref{SFsym}--\eqref{EMsymtl} by a spatial rotation by
$\pi/2$. We find the worldlines $2$, $2'$ to be given by
\selectlcform 
{  
\begin{equation}\label{AccMondSsym}
\begin{aligned}
  \tan\tlt &=
  -\frac{\sqrt{1+a_\oix^2\alpha^2}}{\sinh\bigl(
  \lambda_\dS \alpha^{-1}
  \sqrt{1+a_\oix^2\alpha^2}\bigr)}
  \commae\\
  \tan\tlr &=
  \pm\frac{\cosh\bigl(
  \lambda_\dS \alpha^{-1}
  \sqrt{1+a_\oix^2\alpha^2}
  \bigr)}{a_\oix\alpha}\comma
  \tht,\ph=0\period
\end{aligned}
\end{equation}
The scalar and electromagnetic fields
produced by sources moving along these
worldlines are:
\begin{equation}\label{SFdSBorn}
  \SF_\dSBorn = \frac{\SFq}{4\pi}
  \frac{\sin\tlt}{\sin\tlt+\cos\tlr}\;
  \frac{1}{\retR}\commae
\end{equation}
\begin{widetext}
\begin{equation}\label{EMFdSBorn}
\begin{split}
  \EMF_\dSBorn &= -\frac{\EMq}{4\pi}
  \frac{\alpha^3}{\retR^3}\,
  \frac{1}{(\sin\tlt+\cos\tlr)^3}\;\Bigl[
  - (\sqrt{1+a_\oix^2 \alpha^2} \cos\tlr - a_\oix\alpha\sin\tlt\,)\,
  \cos\tht\,\grad\tlt\wedge\grad\tlr\\
  &\qquad\qquad\quad
  + (\sqrt{1+a_\oix^2 \alpha^2} - a_\oix \alpha \cos\tlr\sin\tlt\,)\,
  \sin\tlr\sin\tht\,\grad\tlt\wedge\grad\tht
  + a_\oix\alpha \sin^2\tlr\cos\tlt\sin\tht\,
  \grad\tlr\wedge\grad\tht
  \Bigr]\commae
\end{split}
\end{equation}
\end{widetext}
where
\begin{equation*}\label{RetR}
  \frac{\retR}{\alpha} = \frac{
  \bigl[(a_\oix\alpha\sin\tlt -
  \sqrt{1+a_\oix^2 \alpha^2}\cos\tlr)^2 +
  \sin^2\tlr\sin^2\tht\bigr]^{\frac12}}
  {\sin\tlt+\cos\tlr}\period
\end{equation*}
}{ 
\begin{equation}\label{AccMondSsym}
\begin{aligned}
  \cot\tlt &=
  -{\sinh\bigl(
  \lambda_\dS \alpha^{-1}
  \sqrt{1+a_\oix^2\alpha^2}\bigr)}/
  \sqrt{1+a_\oix^2\alpha^2}
  \commae\\
  \tan\tlr &=
  \pm{\cosh\bigl(
  \lambda_\dS \alpha^{-1}
  \sqrt{1+a_\oix^2\alpha^2}
  \bigr)}/({a_\oix\alpha})
  \comma
\end{aligned}
\end{equation}
$\tht=0$, $\ph=0$. The scalar and electromagnetic fields are
\begin{gather}
  \SF_\dSBorn = ({\SFq}/{4\pi})\;
  {\sin\tlt}\; ({\sin\tlt+\cos\tlr})^{-1}\;
  {\retR}^{-1}\commae\label{SFdSBorn}\displaybreak[0]\\
\begin{split}
  \EMF_{\dSBorn} &= -\frac{\EMq}{4\pi}
  \frac{\alpha^3}{\retR^3}
  \frac{a_\oix\alpha\sin\tht}{(\sin\tlt+\cos\tlr)^3}\bigl[
  \sin^2\tlr\cos\tlt\,
  \grad\tlr\wedge\grad\tht
  \\&\mspace{2mu}
  - (a_\oix^{-1}\sqrt{a_\oix^{2}+\alpha^{-2}} \cos\tlr - \sin\tlt\mspace{2mu})\,
  \cot\tht\,\grad\tlt\wedge\grad\tlr
  \\&\mspace{2mu}
  + (a_\oix^{-1}\sqrt{a_\oix^{2}+ \alpha^{-2}} - \cos\tlr\sin\tlt\mspace{2mu})\,
  \sin\tlr\,\grad\tlt\wedge\grad\tht
  \bigr]\commae\label{EMFdSBorn}
\end{split}\displaybreak[0]\\
  \frac{\retR}{\alpha} = \frac{
  \bigl[(a_\oix\alpha\sin\tlt -
  \sqrt{1+a_\oix^2 \alpha^2}\cos\tlr)^2 +
  \sin^2\tlr\sin^2\tht\bigr]^{\frac12}}
  {\sin\tlt+\cos\tlr}\period\label{RetR}\nonumber
\end{gather}
}  

In order to understand explicitly the relation of these fields to the
classical Born solutions, consider Minkowski spacetime with spherical
coordinates ${\{\nlt, \nlr, \tht, \ph\}}$ with metric $\mtrc_\nlMink =
- \grad\nlt\formsq + \grad\nlr\formsq + \nlr^2\sphmtrc$.
If we set
\begin{equation*}
  \nlt = -\frac{\alpha \cos\tlt}{\cos\tlr+\sin\tlt}\comma
  \nlr = \frac{\alpha \sin\tlr}{\cos\tlr+\sin\tlt}\commae
\end{equation*}
with $\tht$, $\ph$ unchanged, we find that this Minkowski
space is conformally related to
de~Sitter space as follows (Fig.~\ref{fig:dSMink}):
\begin{equation}\label{dSMinknlCT}
  \mtrc_\dS = \Omega^2\mtrc_\nlMink\comma
  \Omega
  = \frac{\cos\tlr+\sin\tlt}{\sin\tlt}
  = \frac{2\,\alpha^2}{\alpha^2-\nlt^2+\nlr^2}\period
\end{equation}
In coordinates
${\{\nlt,\nlr,\tht,\ph\}}$, which can also be used in %
\selectlcform 
{  
de~Sitter space\footnote{\label{note:confcoor}
  Coordinates $\{\nlt,\nlr,\tht,\ph\}$ differ from the standard
  conformally flat coordinates $\{\hct,\hcr,\tht,\ph\}$
  just by shift in $\tlt$-direction by $\pi/2$
  (cf. Figs.~\ref{fig:dS}, \ref{fig:dSMink}).
  Coordinates $\{\hct,\hcr,\tht,\ph\}$ are related to the
  usual \vague{steady-state} coordinates
  $\{\hceta,\hcr,\tht,\ph\}$ of exponentially expanding
  $k=0$ cosmologies by a simple time rescaling
  ${\hct=-\alpha\exp(-\hceta/\alpha)}$.
  }%
}{ 
de~Sitter space (note~\cite{note:confcoor}),%
}  
the worldlines $2$, $2'$, Eqs.~\eqref{AccMondSsym}, acquire the simple form:
$\tht=0$, $\ph=0$, and
\begin{equation}\label{AccMonMink}
  \nlt = b_\oix \sinh({\lambda_\nlMink}/{b_\oix}) \comma
  \nlr = \pm b_\oix \cosh({\lambda_\nlMink}/{b_\oix})\commae
\end{equation}
where $\lambda_\nlMink$ is the proper time
as measured by $\mtrc_\nlMink$, and%
\selectlcform 
{  
\begin{equation*}
  {b_\oix}/{\alpha} =
  \sqrt{1+a_\oix^2\alpha^2}-a_\oix\alpha\period
\end{equation*}%
}{ 
${b_\oix}/{\alpha} \!=\!
\sqrt{1\!+\!a_\oix^2\alpha^2}-a_\oix\alpha$.%
}  
The worldlines \eqref{AccMonMink} are just two
hyperbolae (Fig.~\ref{fig:dSMink}), representing particles
with uniform acceleration $1/b_\oix$ as measured in
Minkowski space.

Transforming the fields \eqref{SFdSBorn} and
\eqref{EMFdSBorn} into conformally
flat coordinates ${\{\nlt, \nlr, \tht, \ph\}}$, we obtain%
\selectlcform 
{  
\begin{align}
  \SF_\dSBorn &=
  \Omega^{-1}\;\frac{\SFq}{4\pi}\;
  \frac{1}{\retR}\commae\label{SFdSBornnl}\\
\begin{split}
  \EMF_\dSBorn &= -\frac{\EMq}{4\pi}
  \frac{1}{2 b_\oix}\,\frac{\alpha^3}{\retR^3}\;
  \Bigl[
  - 2\, \nlt\,\nlr^2\sin\tht\,
  \grad\nlr\wedge\grad\tht
  \\&\qquad
  - (b_\oix^2+\nlt^2-\nlr^2)\cos\tht\,
  \grad\nlt\wedge\grad\nlr
  \\&\qquad
  + \nlr\, (b_\oix^2+\nlt^2+\nlr^2)\sin\tht\,
  \grad\nlt\wedge\grad\tht
  \Bigr]\commae\label{EMFdSBornnl}
\end{split}
\end{align}%
}{ 
\begin{gather}
  \SF_\dSBorn =
  ({\SFq}/{4\pi})\;\Omega^{-1}\;
  {\retR}^{-1}\commae\label{SFdSBornnl}\\
\begin{split}
  &\EMF_\dSBorn = -\frac{\EMq}{4\pi}
  \frac{\alpha^3}{2 b_\oix}\,\frac{\sin\tht}{\retR^3}\;
  \bigl[
  \nlr\, (b_\oix^2+\nlt^2+\nlr^2)\,
  \grad\nlt\wedge\grad\tht
  \\&\mspace{20mu}
  - (b_\oix^2+\nlt^2-\nlr^2)\cot\tht\,
  \grad\nlt\wedge\grad\nlr
  - 2\, \nlt\,\nlr^2\,
  \grad\nlr\wedge\grad\tht
  \bigr]\commae\label{EMFdSBornnl}
\end{split}
\end{gather}%
}  
the factor $\retR$ now being given by
\begin{equation}
  \retR = \frac{1}{2 b_\oix}
  \sqrt{ (b_\oix^2+\nlt^2-\nlr^2)^2 +
  4\, b_\oix^2\,\nlr^2 \sin^2\tht}\period
\end{equation}

Expressions \eqref{SFdSBorn}, \eqref{EMFdSBorn}, and
\eqref{SFdSBornnl}, \eqref{EMFdSBornnl} represent \emph{the generalized
Born scalar and electromagnetic fields from the sources moving with
constant acceleration $a_\oix$ along the worldlines
\eqref{AccMondSsym}, respectively \eqref{AccMonMink}, in de~Sitter
universe.}

To connect these fields with their counterparts in flat space, note
that they are conformally related by transformation \eqref{dSMinknlCT}.
Under the conformal transformation, the field $\SF_\dSBorn$ in
\eqref{SFdSBornnl} has to be multiplied by factor $\Omega$, which gives
${\SF_\dSBorn = ({\SFq}/{4\pi})\,{\retR}^{-1}}$, and $\EMF_\dSBorn$ in
\eqref{EMFdSBornnl} remains unchanged. The transformed fields then
precisely coincide with the classical Born fields; see e.g.
Refs.~\cite{EriksenGron:2000}, \cite{BicakSchmidt:1989},
\cite{Rohrlich:book}.

In order to see the limit for $\Lambda\rightarrow 0$, we
parameterize the sequence of de~Sitter spaces by $\Lambda$, identifying
them in terms of coordinates ${\{\nlt,\nlr,\tht,\ph\}}$. As
$\Lambda={3/\alpha^2\rightarrow0}$, Eq.~\eqref{dSMinknlCT} implies
${\Omega_\Lambda\rightarrow2}$,
${\mtrc_{\dS\,\Lambda}\rightarrow4\mtrc_\nlMink}$. After the trivial
rescaling of $\nlt$, $\nlr$ by factor $2$, the standard Minkowski
metric is obtained. The limit of the fields \eqref{SFdSBornnl} and
\eqref{EMFdSBornnl}, in which $b_\oix$ is kept constant (cf.
$a_\oix={(1-b_\oix^2\alpha^{-2})/(2b_\oix)}$), leads to the scalar and
electromagnetic Born fields in flat space. Because of the rescaling of
coordinates by factor $2$, we get the physical acceleration
${1/{b_\oix}=2a_\oix}$, and the scalar field rescaled by $1/2$.

What is the character of the generalized Born fields?
Focusing on the electromagnetic case, we first decompose the
field \eqref{EMFdSBorn} into the orthonormal tetrad
${\{\cbv_\mu\}}$ tied to coordinates
${\{\tlt,\tlr,\tht,\ph\}}$; for example,
$\cbv_\tlt={(\alpha^{-1}\sin\tlt\,)\,\cvectil{\tlt}}$, etc., the
dual tetrad $\cbf^\tlt={(\alpha/\sin\tlt\,)\,\grad\tlt}$, etc.
Splitting the field into the electric and magnetic parts,
${F_\dSBorn=\EME\wedge\cbf^\tlt+
\EMB\cdot\cbf^\tlr\wedge\cbf^\tht\wedge\cbf^\ph}$, we get
\begin{equation}\label{EMEBcb}
\begin{aligned}
  \EME&=\frac{\EMq}{4\pi}
  \frac{\alpha\sin^2\tlt}{\retR^3(\sin\tlt+\cos\tlr)^3}
  \times\\&\qquad
  \bigl[
  -\bigl(\sqrt{1+a_\oix^2\alpha^2}\cos\tlr
  -a_\oix\alpha\sin\tlt\bigr)\cos\tht\,\cbv_\tlr
  \\&\qquad
  +\bigl(\sqrt{1+a_\oix^2\alpha^2}
  -a_\oix\alpha\sin\tlt\cos\tlr\bigr)\sin\tht\,\cbv_\tht
  \bigr]\commae\\
  \EMB&=-\frac{\EMq}{4\pi}
  \frac{a_\oix\alpha^2\sin^2\tlt}
  {\retR^3(\sin\tlt+\cos\tlr)^3}
  \cos\tlt\sin\tlr\sin\tht\,\cbv_\ph\period
\end{aligned}
\end{equation}
The fields exhibit some features typical for the
classical Born solution. The toroidal electric field,
$\EME_\ph$, vanishes; only $\EMB_\ph$ is
non-vanishing. At $\tlt=\pi/2$, the moment of time symmetry,
$\EMB_\ph=0$. It vanishes also for $\tht=0$ ---
there is no Poynting flux along the axis of symmetry.

The classical Born field decays rapidly (${\EME\sim r^{-4}}$, ${\EMB\sim
r^{-5}}$) at spatial infinity, but it is \vague{radiative} (${\EME,\EMB}\sim
r^{-1}$) if we expand it along null geodesics ${t-r=\text{constant}}$,
approaching thus null infinity. In de~Sitter spacetime with standard slicing,
the space is finite ($S^3$). However, we can approach infinity along spacelike
hypersurfaces if, for example, we consider the \vague{steady-state} half of
de~Sitter universe (cf. Fig.~\ref{fig:dS}) with flat-space slices, i.e., if we
take the
\vague{conformally flat} time ${\hct=\text{constant}}$%
\selectlcform 
{  
(see footnote~\ref{note:confcoor}).%
}{ 
(note \cite{note:confcoor}).%
}  
Introducing the orthogonal tetrad tied to conformally flat coordinates
$\{\hct,\hcr,\tht,\ph\}$, the tetrad components of the fields decay as
$\hcr^{-2}$ at ${\hct=\text{constant}}$, ${\hcr\rightarrow\infty}$, so that the
Poynting flux falls off as $\hcr^{-4}$.

The fields decay very rapidly along \emph{timelike} worldlines as
$\scry^+$ is approached. This is caused by the exponential expansion of
slices ${\tau=\text{constant}}$ (cf. Eq.~\eqref{dSMtrctauchi}). As
${\tau\rightarrow\infty}$ the electric field \eqref{EMEBcb} becomes
radial, ${\EME_\tlr\sim\exp(-2\tau/\alpha)}$, and
${\EMB_\ph\sim\exp(-2\tau/\alpha)}$. The energy density,
${\EMU=\frac12(\EME^2+\EMB^2)}$, decays as ${(\text{expansion
factor})^{-4}}$
--- as energy density in the radiation dominated standard cosmologies. The
density of the conserved energy
  ${\EMU_\conf=(\alpha/\sin\tlt\,)\EMU}\sim{\exp(-3\tau/\alpha)}$
(determined by a timelike conformal Killing vector
${\cvectil{\tlt}}\,$) gets rarified at the same rate that the volume
increases.

\begin{figure}
\includegraphics{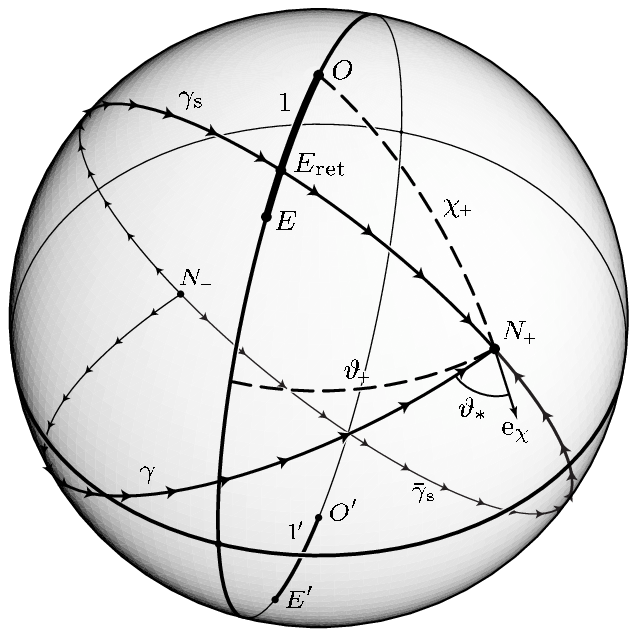}
\caption{\label{fig:spacesec}
\selectlcform 
{  
Space trajectories of null geodesics $\gamma$, $\gamma_{\mathrm{s}}$ and
$\bar\gamma_{\mathrm{s}}$ indicated on the slice ${\tlt=\text{constant}}$
(${\ph=0}$). The slice contracts from infinite volume to its minimal size at
${\tlt=\pi/2}$ and then expands again. Charges $1$, $1'$ move along the axis
$\tht=0$ from poles $O$, $O'$ to points $E$, $E'$ and back. The geodesics
$\gamma$, $\gamma_{\mathrm{s}}$ and $\bar\gamma_{\mathrm{s}}$
start at point $N_\mix$ at ${\tlt=0}$ and arrive at
point $N_\pix$ (with coordinates $\chp$, $\thp$) at $\tlt=\pi$. The direction
of the geodesic $\gamma$ at $N_\pix$ is specified by angles $\thst$, $\phst$
($\phst$ describes rotation around $\cbv_\chi$ in the dimension not seen).
The geodesic $\gamma_{\mathrm{s}}$ crosses the worldline of particle $1$ at
$E_{\mathrm{ret}}$, $\bar\gamma_{\mathrm{s}}$ reaches $N_\pix$ from the
opposite direction.%
}{ 
Space trajectories of null geodesics $\gamma$, $\gamma_{\mathrm{s}}$ and
$\bar\gamma_{\mathrm{s}}$ indicated on the slice ${\tlt=\text{constant}}$
(${\ph=0}$). Charges $1$, $1'$ move along $\tht=0$ from poles $O$, $O'$ to
points $E$, $E'$ and back. $\gamma$, $\gamma_{\mathrm{s}}$ and
$\bar\gamma_{\mathrm{s}}$ start at $N_\mix$ at ${\tlt=0}$ and arrive at
$N_\pix$ (with coordinates $\chp$, $\thp$) at $\tlt=\pi$. The direction of
$\gamma$ at $N_\pix$ is specified by angles $\thst$, $\phst$ ($\phst$ describes
rotation around $\cbv_\chi$ in the dimension not seen). $\gamma_{\mathrm{s}}$
crosses the worldline of particle $1$ at $E_{\mathrm{ret}}$;
$\bar\gamma_{\mathrm{s}}$~reaches $N_\pix$ from the opposite direction.%
}  
}%
\end{figure}

Will a slower decay occur if $\scry^+$ is approached along null
geodesics? To study the asymptotic behavior of a field along a null
geodesic (see, e.g., Ref.~\cite{PenroseRindler:book}), we have to (i)
find a geodesic and parameterize it by an affine parameter $\afp$, (ii)
construct a tetrad parallelly propagated along the geodesic, and (iii)
study
the asymptotic expansion of the tetrad components of the field.%
\selectlcform 
{  
The details of these calculations will be presented elsewhere. Here
we just list some typical conclusions. Along%
}{ 
We find that along%
}  
null geodesics lying in the axis $\tht=0$ (thus crossing the particles'
worldlines) the \vague{radiation field}, i.e., the coefficient of the
leading term in ${1/\afp}$, vanishes, as could have been anticipated
--- particles do not radiate in the direction of their acceleration.
The radiation field also vanishes along null geodesics reaching
infinity along directions \emph{opposite} to those of geodesics
emanating from the particles (see Fig.~\ref{fig:spacesec}). Along all
other geodesics, the field \emph{has} radiative character. Along a null
geodesic coming from a general direction to a general point on
$\scry^+$, we find the electric and magnetic fields (in a parallelly
transported tetrad~${\{\pbv_\mu\}}$) to be perpendicular one to the
other, equal in magnitude, and proportional to $\afp^{-1}$. The
magnitude of Poynting flux,
${\abs{\EMP_{(\pbv)}}}={\abs{\EME_{(\pbv)}}{}^{\!2}}={\abs{\EMB_{(\pbv)}}{}^{\!2}}$,
is%
\selectlcform 
{  
\begin{equation}
\begin{split}
  &\abs{\EMP_{(\pbv)}} =
  \frac{\EMq^2}{(4\pi)^2}\,
  \frac1{4\alpha^2\,(1+a_\oix^{2}\alpha^2\cos^2\thp)^{3}\sin^4\chp}
  \frac1{\afp^2}
  \\&\qquad
  \bigl[
  \bigl(a_\oix\alpha\sin\thp\cos\phst + \sqrt{1+a_\oix^{2}\alpha^2}\sin\thst\bigr)^2
  \\&\mspace{160mu}
  +\bigl(a_\oix\alpha\sin\thp\cos\thst\sin\phst\bigr)^2
  \bigr]
\end{split}\raisetag{34pt}
\end{equation}%
}{ 
\begin{equation}
\begin{split}
  &\abs{\EMP_{(\pbv)}} \!=
  \frac{\EMq^2}{(4\pi)^2}
  \frac{a_\oix^2\sin^2\thp\csc^4\chp}{4(1+a_\oix^{2}\alpha^2\cos^2\thp)^{3}}
  \bigl[
  \cos^2\thst\sin^2\phst
  \\&\mspace{64mu}
  +\bigl(\cos\phst +
  a_\oix^{-1}\sqrt{a_\oix^{2}+\alpha^{-2}}
  \sin\thst\csc\thp\bigr)^2
  \bigr]\;\afp^{-2}
\end{split}
\end{equation}%
}  
(see Fig.~\ref{fig:spacesec} for the definition of angles $\chp$,
$\thp$, $\thst$, $\phst$). These results are typical for a
\emph{radiative} field. Most interestingly, this radiative aspect
depends on the specific geodesic along which a given point on
\emph{spacelike} $\scry^+$ is approached
(cf.~\cite{PenroseRindler:book}). Moreover, the radiative character
does not disappear even for static sources but it does along null
geodesics emanating from such sources.

Since the field can be interpreted as the combination of retarded and advanced
effects, similarly to the flat space case \cite{BicakSchmidt:1989}, the
radiation reaction force \cite{DeWittBrehme:1960} also vanishes.

In summary, we have constructed the fields of uniformly accelerated charges in
a de Sitter universe which go over to classical Born fields in the limit
${\Lambda\rightarrow 0}$. Aside from some similarities found, the generalized
fields provide the models showing how a positive cosmological constant implies
essential differences from physics in flat spacetime: Advanced effects occur
inevitably, and the character of the far fields depends substantially on the
way in which future (spacelike) infinity is approached.
Since vacuum energy seems to be dominant in the universe, it is of
interest to understand fundamental physics in the vacuum dominated de
Sitter spacetime.

\begin{acknowledgments}
The authors thank the Albert-Einstein-Institute, Golm, for hospitality, and
Jeff Winicour for very helpful comments on the manuscript. We have been
supported by the grants GA\v{C}R 202/99/026 and GAUK 141/2000.
\end{acknowledgments}



\end{document}